\documentclass[amsmath,amssymb,nofootinbib,twocolumn]{revtex4}
\usepackage{graphicx}
\usepackage{dcolumn}
\usepackage{color}
\usepackage{scrextend}
\usepackage{subfig}
\usepackage{bm}
\usepackage{amsmath,amssymb,graphicx}
\usepackage{epsfig}
\usepackage{enumerate}
\usepackage{array}
\usepackage{multirow}

\begin{document}
\title
{Dirac-isotonic oscillators in \((1+1)\) and \((2+1)\) dimensions}
\author{Aritra Ghosh$^1$\footnote{aritraghosh500@gmail.com, ag34@iitbbs.ac.in} and Bhabani Prasad Mandal$^2$\footnote{bhabani.mandal@gmail.com, bhabani@bhu.ac.in}}
\affiliation{$^{1}$School of Basic Sciences, Indian Institute of Technology Bhubaneswar, Jatni, Khurda, Odisha 752050, India\\
$^2$ Department of Physics, Banaras Hindu University, Varanasi, Uttar Pradesh 221005, India}
\vskip-2.8cm
\date{\today}
\vskip-0.9cm

\vspace{5mm}

\textit{Dedicated to our close collaborator, Professor Bijan Bagchi,\\ on his 75th birthday with deep respect and admiration}

\vspace{5mm}
\begin{abstract}
We discuss the Dirac oscillator in \((1+1)\) and \((2+1)\) dimensions and generalize it in the spirit of the isotonic oscillator using supersymmetric quantum mechanics. In \((1+1)\) dimensions, the Dirac oscillator returns to the quantum harmonic oscillator in the non-relativistic limit, while its generalization maps to the isotonic oscillator. We describe exact solutions of these generalized systems and also present their non-relativistic limits. Finally, based on supersymmetric quantum mechanics, we show that a generalized Dirac oscillator in \((2+1)\) dimensions can be mapped to an anti-Jaynes-Cummings-like Hamiltonian in which the spin operators couple with the supercharges. 
\end{abstract}

\maketitle

\section{Introduction}

The Dirac equation (see for example, \cite{dirac}) is one of the most important discoveries of the 20th century. It has been solved exactly in multiple cases, including the free particle and the hydrogen atom. Further, it has appeared frequently in recent literature, mainly in the context of graphene as the (electronic) dispersion relation of the latter can take a form linear in the momentum \cite{graphene}. It\^{o} and coworkers \cite{diracosc1} studied the Dirac equation by adding a linear term $-imc\omega\beta{{\mbox{\boldmath $\alpha $}}}{\bf \cdot r}$ and subsequently it was observed that in the non-relativistic limit, the system reduces to a harmonic oscillator with a strong spin-orbit coupling and so referred to as the Dirac oscillator by  Moshinsky and Szczepaniak \cite{diracosc} (see also, \cite{diracoscquesne}). The Dirac oscillator has been studied extensively from different viewpoints including supersymmetric quantum mechanics \cite{susyquesne,susydo} (see also, \cite{susydirac,bagchidirac}), magnetism \cite{diracB,app3}, and quantum phase transitions \cite{qpt, qpt1}.
Moreover, the Dirac oscillator has attracted a lot of attention and has found many physical applications in various areas of physics \cite{app7,app6,app10,app1,app4,app8,app9,app2,app5,ext1,silva,ext2}. Various other exact results concerning the \((2+1)\)-dimensional Dirac equation have been reported in the literature (see for example, \cite{2d1,2d2,2d3,2d4}). An exciting property of the Dirac oscillator lies in its connection with quantum optics \cite{opt,opt1,opt2}, where in \((2+1)\) dimensions it maps to the Jaynes-Cummings/anti-Jaynes-Cummings models \cite{qpt1,diracB} describing the atomic transitions in a two-level system. 

\vspace{2mm}

The harmonic oscillator plays a key role in classical and quantum mechanics, mainly due to the fact that it is an exactly-solvable system but also due to its ubiquitous appearance in different problems of physical interest, from mechanics to quantum field theory. An intriguing aspect of the classical harmonic oscillator is the insensitivity of the period of oscillation to the amplitude or energy, i.e., the system admits a constant frequency, irrespective of the maximum displacement from the origin. Such systems are called `isochronous' and have been studied in detail by several authors \cite{urabe,CV,Calo,PG_AGC_symp,PG_AGC_iso}; particularly notable amongst these is the work of Urabe \cite{urabe}. It has been demonstrated that for one-dimensional systems admitting potentials that are rational functions of the position variable, there are only two potentials that support isochronous oscillations \cite{CV}: 
\begin{equation}\label{1disochronous}
V_{\rm ho}(x) = \frac{m \omega^2 x^2}{2}, \quad \quad V_{\rm io}(x) = A x^2 + \frac{B}{x^2}, 
\end{equation} where \(m, \omega, A, B > 0\) are constants. While \(V_{\rm ho}(x)\) which is the harmonic potential is defined on the entire real line, the other one, i.e., \(V_{\rm io}(x)\), dubbed the isotonic potential is defined on the half-line, i.e., either for \(x > 0\) or for \(x < 0\) due to the quadratic singularity at \(x = 0\); for definiteness, we will take \(x > 0\). Various other isochronous systems can often be mapped to either the harmonic oscillator or the isotonic oscillator (see for example, \cite{Lie1,Lie2,mom}). Therefore, apart from the harmonic oscillator whose importance is well known, the isotonic oscillator presents yet another intriguing exactly-solvable problem with properties similar to the former. The classical (Newtonian) equation of motion arising from such a potential is well known as the Ermakov-Pinney equation \cite{EP} which arises in the context of the Ermakov-Lewis invariant of the time-dependent harmonic oscillator (see \cite{paddy} and references therein). Although the importance of the harmonic potential is well appreciated, the isotonic potential is relatively less discussed despite its remarkable properties. Physically, a system governed by the isotonic potential corresponds to the dynamics of the radial variable \(r = \sqrt{x^2 + y^2}\) of a two-dimensional isotropic oscillator with the potential \(V(x) \propto (x^2 + y^2)\) \cite{PG_AGC_iso} and is of great interest in quantum optics \cite{Opt_isotonic} and in the theory of coherent states \cite{coherent_states1,coherent_states2}.

\vspace{2mm}

Quantum mechanically, isochronous systems are associated with an equispaced spectra \cite{iso_car}. In fact, the isotonic oscillator is exactly solvable and admits a spectrum of equispaced energy levels \cite{iso1,iso2}. Due to the importance of the isotonic potential as asserted above, it seems worthwhile to explore the extension of the Dirac oscillator in the spirit of the isotonic potential. One must, however, mention some preceding developments \cite{dirac_iso_old_0,dirac_iso_old} in which the Dirac equation has been analyzed in the context of the isotonic potential. Nevertheless, the present approach rests on supersymmetric quantum mechanics (see also, \cite{bagchidirac}) and is distinct from the preceding developments on the relativistic generalization of the isotonic oscillator. Although some preliminary results for the same in \((1+1)\) dimensions appear in \cite{bagchidirac}, in this paper, we will propose exactly-solvable extensions of the Dirac equation in the spirit of the isotonic oscillator in both \((1+1)\) and \((2+1)\) dimensions and will also treat their harmonic counterparts. The remarkable feature that we will show is that the so-called Dirac-isotonic oscillators in \((1+1)\) and \((2+1)\) dimensions both admit an equispaced spectrum of levels for \(E^2\) (squared energy eigenvalues). Interestingly, for the \((2+1)\)-dimensional extension, the non-relativistic limit admits an intriguing coupling between the angular momentum and the centrifugal term. Finally, based on supersymmetric quantum mechanics and the known connection between the Dirac oscillator and the Hamiltonian of a two-level system interacting with a quantized single-mode optical field \cite{qpt1,diracB}, we shall present a generalization of the same in which the two-level system couples with the supercharge operators. With this brief introduction, let us begin our formal analysis with the \((1+1)\)-dimensional case; the \((2+1)\)-dimensional case will be taken up subsequently.

\section{(1+1)-dimensional case}
Let us begin with the \((1+1)\)-dimensional Dirac Hamiltonian \cite{dirac}
\begin{equation}
H = c \alpha (p_x - i \beta W(x)) + \beta m c^2,
\end{equation} where \(\alpha\) and \(\beta\) are 2 \(\times\) 2 matrices, \(W(x)\) is a real-valued function that we will specify below, and other symbols have their usual meanings. For our purposes, we have \(\alpha = \sigma_x\) and \(\beta = \sigma_z\), where \(\sigma_x\), \(\sigma_y\), and \(\sigma_z\) are the Pauli matrices. Thus, in the matrix form, the Hamiltonian turns out to be
\begin{equation}
H =
\begin{pmatrix}
mc^2 & c(p_x + i W(x)) \\
c(p_x - i W(x)) & -mc^2
\end{pmatrix} ,
\end{equation} and which is obviously Hermitian for a real-valued function \(W(x)\). Taking
\begin{equation}\label{diracwave}
\Psi = \begin{pmatrix}
\psi_1 \\
\psi_2
\end{pmatrix} 
\end{equation} to be a time-independent eigenfunction of \(H\) (the Dirac wavefunction), using \(H \Psi = E \Psi\) for real \(E\), one gets the following first-order equations:
\begin{eqnarray}
c(p_x - i W(x)) \psi_1  &=& (E + mc^2) \psi_2, \label{firstorder1}\\
c(p_x + i W(x)) \psi_2  &=& (E - mc^2) \psi_1. \label{firstorder2}
\end{eqnarray}
Considering (\ref{firstorder2}), multiplying both sides with \((E + mc^2)\) and upon using (\ref{firstorder1}), we get (we will set \(\hbar=1\))
\begin{equation}\label{RelTISE2}
c^2 \bigg[ - \frac{d^2}{dx^2} + W(x)^2 - W'(x) \bigg] \psi_1 = [E^2 - (mc^2)^2] \psi_1,
\end{equation}
where we have used \(p = - i (d/dx)\). Writing \(E = mc^2 + \epsilon\), in the non-relativistic limit, i.e., for \(c \rightarrow \infty\), we find \(E^2 - (mc^2)^2 \approx 2mc^2 \epsilon\), and this means equation (\ref{RelTISE2}) turns out to be 
\begin{equation}\label{nonRelTISE}
 \bigg[ - \frac{1}{2m} \frac{d^2}{dx^2} + \frac{W(x)^2}{2m} - \frac{ W'(x)}{2m} \bigg] \psi_{1} =  \epsilon \psi_{1}.
\end{equation} One can similarly derive the evolution equation for \(\psi_2\) which makes the supersymmetric connection between \(\psi_1\) and \(\psi_2\) clear if we identify \(W(x)\) as the superpotential \cite{susydirac,bagchidirac}. However, throughout this paper we will explicitly consider only \(\psi_1\) for the sake of simplicity as \(\psi_2\) can then be obtained from equation (\ref{firstorder1}) once \(\psi_1\) is known. Thus, in the following, we will refer to \(\psi_1\) as the wavefunction. 

\subsection{(1+1)-dimensional Dirac oscillator}
In order to observe the \((1+1)\)-dimensional Dirac oscillator, the counterpart of the case studied in \cite{diracosc} (see also, \cite{diracB}), let us take
\begin{equation}\label{harmopot}
W(x) = m \omega x, \quad \quad x \in \mathbb{R},
\end{equation} where \(m, \omega > 0\) are constants. Thus, equation (\ref{RelTISE2}) turns out to be
\begin{equation}\label{RelTISE1dosc}
 \bigg[ - \frac{d^2}{dx^2} + m^2 \omega^2 x^2 - m \omega \bigg] \psi_1 = \frac{[E^2 - (mc^2)^2]}{c^2} \psi_1.
\end{equation}
The above equation possesses the same operator structure as the time-independent Schr\"odinger equation of a harmonic oscillator. In particular, if we write 
\begin{equation}\label{RelTISE1dosc2}
m \omega \bigg[ - \frac{d^2}{du^2} + u^2 \bigg] \psi_1 = \frac{[E^2 - (mc^2)^2 + m c^2 \omega]}{c^2} \psi_1,
\end{equation} where we have defined \(u = \sqrt{m \omega} x\), then the solution of the above-mentioned equation is
\begin{widetext}
\begin{equation}\label{1doscsol}
\psi_{1,n}(x) = \frac{\pi^{1/4}}{\sqrt{2^n n!}} e^{-m \omega x^2/2} H_n(\sqrt{m \omega}x),\quad \quad E_n^2 = (mc^2)^2 \bigg[1 + \frac{2n \omega}{mc^2}\bigg],
\end{equation} where \(n = 0,1,2,\cdots\) and \(\{H_n(\cdot)\}\) are the Hermite polynomials.
\end{widetext} This is the \((1+1)\)-dimensional analogue of the Dirac oscillator described by Moshinsky and Szczepaniak \cite{diracosc}. In the non-relativistic limit, equation (\ref{nonRelTISE}) becomes
\begin{equation}\label{nonRelTISE1}
 \bigg[ - \frac{1}{2m} \frac{d^2}{dx^2} + \frac{m \omega^2 x^2}{2} - \frac{\omega}{2} \bigg] \psi_{1} =  \epsilon \psi_{1},
\end{equation}
which is just the harmonic oscillator and can be solved to give the spectrum \(\epsilon_n = n \omega\). Note that unlike Moshinsky and Szczepaniak \cite{diracosc}, the non-relativistic limit does not contain a spin-orbit term -- this is due to the fact that we have considered the \((1+1)\)-dimensional Dirac oscillator unlike the \((3+1)\)-dimensional case as considered in \cite{diracosc}. 

\begin{widetext}
\subsection{(1+1)-dimensional Dirac-isotonic oscillator}
In what follows, we will extend this construction to describe the relativistic analogue of the isotonic oscillator \cite{iso1,iso2}. For that, let us consider the following form of the function \(W(x)\): 
\begin{equation}\label{superpotisotonic}
W(x) = a x + \frac{b}{x}, \quad \quad x > 0, 
\end{equation} where \(a, b > 0\) are constants. Thus, the problem is restricted to the real half-line for which \(x > 0\), i.e., the superpotential is singular at \(x = 0\) (see also, \cite{bagchidirac,susypani}). The limit \(b \rightarrow 0\) recovers the harmonic-oscillator case (\(a = m \omega\)) but only on the real half-line. Now, substituting the superpotential (\ref{superpotisotonic}) into equation (\ref{RelTISE2}), one gets the following time-independent Klein-Gordon equation for \(\psi_1\):
\begin{eqnarray}
 \bigg[ - \frac{d^2}{dx^2} + a^2 x^2 + \frac{b(b+ 1)}{x^2} \bigg] \psi_{1} = \frac{[E^2 - (mc^2)^2 + ac^2(-2b + 1)]}{c^2} \psi_{1}. \label{1ddiracisotonic}
\end{eqnarray}  We will say that equation (\ref{1ddiracisotonic}) represents a `Dirac-isotonic' oscillator, defined on \(x > 0\). Since the left-hand side has the same operator structure as the Schr\"odinger equation, we can easily solve the above-mentioned equation following the treatment presented in \cite{iso1}. The wavefunctions turn out to be (up to normalization factors)
\begin{equation}
\psi_{1,n}(x) \sim x^\nu e^{-\frac{ax^2}{2}} {_1}F_1\bigg(-n, \nu + \frac{1}{2}, ax^2 \bigg), \quad \quad \nu = \frac{1}{2} [\sqrt{1 + 4b(b+1)} + 1], 
\end{equation} while the spectrum is
\begin{equation}
E_n^2 = (mc^2)^2 \bigg[ 1 + \frac{a}{m^2 c^2} \big(4n +2b + 1 + \sqrt{1 + 4 b(b+1)}\big) \bigg],
\end{equation} where \(n=0,1,2,\cdots\).
 The function \({_1}F_1(\cdot, \cdot, \cdot)\) is the confluent hypergeometric function that solves Kummer's differential equation and because \(n\) is a positive integer, it can be expressed in terms of Laguerre polynomials \cite{handbook}. In the non-relativistic limit, the equation (\ref{1ddiracisotonic}) returns to a Schr\"odinger equation of a particle with mass \(m\) in an isotonic potential (with constant shift):
\begin{eqnarray}
 \bigg[ - \frac{1}{2m} \frac{d^2}{dx^2} + \frac{a^2 x^2}{2m} + \frac{b(b+ 1)}{2m x^2} + \frac{a(2b-1)}{2m} \bigg] \psi_{1} = \epsilon \psi_{1}, \label{1ddiracisotonicnonrel}
\end{eqnarray} where \(\epsilon = E - mc^2\). Notice that upon putting \(a = m \omega\) and \(b = 0\), the above equation reduces exactly to equation (\ref{nonRelTISE1}), as anticipated. In this light, one can interpret equation (\ref{1ddiracisotonicnonrel}) to be the Schr\"odinger equation for a particle of mass \(m\) as confined within a one-dimensional isotonic potential with the energy shift \(\frac{a(2b-1)}{2m}\). As with the case of equation (\ref{nonRelTISE1}), there is no spin-orbit term because the definition of orbital angular momentum requires more than one spatial dimension. In what follows, we will explore the (2+1)-dimensional case.

\section{(2+1)-dimensional case}
Let us now consider the \((2+1)\)-dimensional Dirac Hamiltonian \cite{dirac}
\begin{equation}
H = c \alpha_x (p_x - i \beta W_x(\mathbf{r})) + c \alpha_y (p_y - i \beta W_y(\mathbf{r})) + \beta m c^2,
\end{equation} where \(\mathbf{r} = (x,y)\), \(\alpha_x = \sigma_x\), \(\alpha_y = \sigma_y\), and \(\beta = \sigma_z\). The `superpotential' is a vector-valued function of \(\mathbf{r} = (x,y)\) with \(x\) and \(y\) components being given by \(W_x=W_x(\mathbf{r})\) and \(W_y=W_y(\mathbf{r})\), respectively. These are real-valued functions. Explicitly, one can write
\begin{equation}\label{2dgenmatrixH}
H =
\begin{pmatrix}
mc^2 & cP^\dagger + i c \mathcal{W}^\dagger(\mathbf{r}) \\
cP - i c \mathcal{W}(\mathbf{r}) & -mc^2
\end{pmatrix} ,
\end{equation} where \(P = p_x + i p_y\) and \(\mathcal{W}(\mathbf{r}) = W_x(\mathbf{r}) + i W_y(\mathbf{r})\). Taking the Dirac wavefunction to be of the form dictated by (\ref{diracwave}) with \(\psi_{1,2} = \psi_{1,2}(x,y)\), we get 
\begin{eqnarray}
 c(P - i\mathcal{W}(\mathbf{r})) \psi_1 &=& (E + mc^2)\psi_2, \\
 c(P^\dagger + i\mathcal{W}^\dagger(\mathbf{r})) \psi_2 &=& (E - mc^2)\psi_1.
\end{eqnarray}
Focusing on \(\psi_1\), combining the two first-order equations above, one finds the following second-order equation: 
\begin{eqnarray}\label{2ddirac}
 c^2 \big[ p_x^2 + p_y^2 + W_x^2 + W_y^2 + 2(W_y p_x - W_x p_y) - i ((p_x W_x) + (p_y W_y)) + ((p_xW_y) - (p_y W_x)) \big] \psi_1 =[E^2 - (mc^2)^2]\psi_1. 
\end{eqnarray}
Let us discuss two cases now. 

\subsection{(2+1)-dimensional Dirac oscillator}
Taking \(W_x = m \omega x\) and \(W_y = m \omega y\), one finds \((p_xW_y) = 0\) and \((p_y W_x) = 0\). Thus, one has
\begin{eqnarray}
  \bigg[ - \bigg( \frac{\partial^2}{\partial x^2} + \frac{\partial^2}{\partial y^2}\bigg) + m^2 \omega^2 (x^2 + y^2) - 2 m \omega L_z - 2 m \omega \bigg] \psi_1 = \frac{[E^2 - (mc^2)^2]}{c^2}\psi_1,\label{abcdef111}
\end{eqnarray} where \(L_z = x p_y -y p_x\). This is the \((2+1)\)-dimensional analogue of the Dirac oscillator \cite{diracosc}. In order to solve it, let us resort to polar coordinates \((r,\theta)\); taking the ansatz \(\psi_1(r,\theta) = e^{i m_l \theta} \phi(r)\), we get
\begin{equation}
- \frac{d^2 \phi}{dr^2} - \frac{1}{r} \frac{d\phi}{dr} + \bigg[ m^2 \omega^2 r^2 + \frac{m_l^2}{r^2} \bigg] \phi =  \frac{[E^2 - (mc^2)^2 + 2 mc^2 \omega + 2 m c^2 m_l \omega]}{c^2}\phi,
\end{equation}
which can be solved using the known techniques. Defining \(\Lambda = \frac{[E^2 - (mc^2)^2 + 2 mc^2 \omega + 2m c^2 m_l \omega]}{c^2}\) and \(\chi(r) = \sqrt{r} \phi(r)\), one gets the simplified equation
\begin{equation}\label{chi1}
- \frac{d^2 \chi}{dr^2} + \bigg[ m^2 \omega^2 r^2 + \frac{m_l^2 - \frac{1}{4}}{r^2} \bigg] \chi =  \Lambda \chi. 
\end{equation} This looks like equation (\ref{1ddiracisotonic}) and can be solved exactly for \(m_l^2 \geq 1/4\) to give (up to normalization factors)
\begin{equation}\label{harmsol11}
\chi_n(r) \sim r^{\left(|m_l| + \frac{1}{2}\right)} e^{- \frac{m \omega r^2}{2}} {_1}F_1\bigg(-n, |m_l| + 1, m \omega r^2\bigg), \quad \quad E_n^2 = (mc^2)^2 \bigg[1 + \frac{2 \omega}{mc^2} (2n + |m_l| - m_l)\bigg], 
\end{equation} where \(n = 0,1,2,\cdots\). It may be remarked that the spectrum obtained above is identical to that found in \cite{silva} (see also, \cite{app6}). Notice that the full time-independent wavefunctions are obtained as \(\psi_{1,n}(r,\theta) = r^{-1/2} e^{im_l \theta} \chi_n(r)\). In the non-relativistic limit (\(c\rightarrow \infty\)), \(E^2 - (mc^2)^2 \approx 2 m c^2 \epsilon\), meaning that one obtains the following Schr\"odinger equation:
\begin{equation}\label{nonrel21dirac}
 \bigg[ - \frac{1}{2m} \bigg( \frac{\partial^2}{\partial x^2} + \frac{\partial^2}{\partial y^2}\bigg) + \frac{m \omega^2}{2} (x^2 + y^2) - \omega L_z - \omega \bigg] \psi_1 = \epsilon \psi_1,
\end{equation} and the spectrum simply reads \(\epsilon_n = \omega \big(2n - m_l \big)\). Thus, in the non-relativistic limit, the problem reduces to the two-dimensional harmonic oscillator but with a spin-orbit term that contributes \(-\omega m_l\) to the spectrum.

\subsection{A nonlinear superpotential}
Inspired from the choice of superpotential (\ref{superpotisotonic}), let us take \(\mathbf{W}(\mathbf{r}) \equiv W_x (\mathbf{r}) \hat{x} + W_y (\mathbf{r}) \hat{y} = a \mathbf{r} + \frac{b \mathbf{r}}{|\mathbf{r}|^2}\). This implies that
\begin{equation}
W_x = a x + \frac{b x}{x^2 + y^2}, \quad \quad W_y = a y + \frac{b y}{x^2 + y^2}.
\end{equation}  Thus, equation (\ref{2ddirac}) becomes 
\begin{eqnarray}\label{021general}
\bigg[ - \bigg( \frac{\partial^2}{\partial x^2} + \frac{\partial^2}{\partial y^2}\bigg) + a^2 (x^2 + y^2) + \frac{b^2}{x^2 + y^2} - 2 \bigg(a + \frac{b}{x^2 + y^2}\bigg) L_z  \bigg] \psi_1 = \frac{[E^2 - (mc^2)^2 - 2ac^2(b-1)]}{c^2}\psi_1,
\end{eqnarray} which matches with equation (\ref{abcdef111}) for \(a = m \omega\) and \(b = 0\). It is convenient to employ polar coordinates \((r,\theta)\) on the plane using which, together with the ansatz \(\psi_1(r,\theta) = e^{i m_l \theta} \phi(r)\), one finds
\begin{eqnarray}\label{21general}
 - \frac{d^2 \phi}{d r^2} - \frac{1}{r} \frac{d \phi}{d r} + a^2 r^2 \phi + \frac{b^2 + m_l(m_l - 2 b)}{r^2} \phi  = \frac{[E^2 - (mc^2)^2 + 2 a m_l c^2 - 2ac^2(b-1)]}{c^2}\phi.
\end{eqnarray}
The term with the first derivative can be eliminated by putting \(\chi(r) = \sqrt{r} \phi(r)\) and one then finds
\begin{eqnarray}\label{diracosc2dradialtosolve}
 - \frac{d^2 \chi}{d r^2} + \bigg[a^2 r^2 + \frac{(m_l - b)^2 - \frac{1}{4}}{r^2} \bigg]\chi  = \Gamma \chi,
\end{eqnarray} where we have identified \(\Gamma = \frac{[E^2 - (mc^2)^2 + 2a m_l c^2 - 2ac^2(b-1)]}{c^2}\). The above-mentioned equation has the same operator structure as equations (\ref{1ddiracisotonic}) and (\ref{chi1}), and can be solved exactly for \((m_l - b)^2 \geq 1/4\) to give
\begin{equation}\label{gensol1}
\chi_n(r) = r^{|m_l - b| + \frac{1}{2}} e^{- \frac{ar^2}{2}} {_1}F_1 \big(-n, |m_l - b| + 1, ar^2 \big), \quad \quad E_n^2 = (mc^2)^2 \bigg[1 + \frac{2a}{m^2 c^2} [2n + |m_l - b| - (m_l - b)] \bigg],
\end{equation} where \(n = 0,1,2,\cdots\). Taking \(a = m \omega\) and \(b = 0\) gives us the results (\ref{harmsol11}); the effect of \(b\) is therefore to shift the quantum number \(m_l\) as is also evident from equation (\ref{diracosc2dradialtosolve}), i.e., we can alternatively define \((m_l)_{\rm eff} = m_l - b\) such that the results (\ref{gensol1}) are obtained by replacing \(m_l\) with \((m_l)_{\rm eff}\) in the results (\ref{harmsol11}). Notice that the full time-independent wavefunctions are obtained as \(\psi_{1,n}(r,\theta) = r^{-1/2} e^{im_l \theta} \chi_n(r)\). Finally, let us remark that the non-relativistic limit is obtained straightforwardly; in particular, in this limit, equation (\ref{21general}) becomes 
\begin{equation}
 \bigg[ -\frac{1}{2m} \bigg( \frac{d^2}{d r^2} + \frac{1}{r} \frac{d}{d r} \bigg) + \frac{a^2 r^2}{2m} + \frac{(m_l - b)^2}{2mr^2} - \frac{a m_l}{m} + \frac{a(b-1)}{m} \bigg]\phi   = \epsilon \phi,
\end{equation} which resembles the isotonic oscillator for the radial variable. Notice that upon putting \(a = m \omega\) and \(b = 0\), one can exactly recover the (radial) equation representing the non-relativistic limit of the Dirac oscillator. A few comments are in order. One can re-write the above-mentioned equation as 
\begin{equation}\label{nonrelrewrite00}
\bigg[ \hat{T} + \hat{V} - \frac{a m_l}{m} + \frac{a(b-1)}{m} \bigg]\phi   = \epsilon \phi,
\end{equation}
 where the kinetic-energy and potential-energy operators acting on the radial part of the wavefunction are
 \begin{equation}
 \hat{T} = -\frac{1}{2m} \bigg( \frac{d^2}{d r^2} + \frac{1}{r} \frac{d}{d r} \bigg)  + \frac{m_l^2}{2mr^2}, \quad \quad \hat{V} = \frac{a^2 r^2}{2m}  + \frac{b^2}{2mr^2} - \frac{m_l b}{m r^2}.
 \end{equation} Notice that the potential energy is dependent on the quantum number \(m_l\) which couples with the centrifugal term, a novel feature that is associated with the \((2+1)\)-dimensional Dirac-isotonic oscillator; this coupling vanishes upon taking \(b \rightarrow 0\) in which case one recovers the non-relativistic limit of the Dirac oscillator of Moshinsky and Szczepaniak by choosing \(a = m \omega\) as given in equation (\ref{nonrel21dirac}). The coupling between the parameters \(m_l\) and \(b\) manifests by causing a shift in the quantum number \(m_l\) in the manner \(m_l \rightarrow m_l - b\). In equation (\ref{nonrelrewrite00}), the term \(- \frac{a m_l}{m}\) is the spin-orbit term while \(\frac{a(b-1)}{m} \) provides a constant shift of the spectrum.
 \end{widetext}
 
 \section{Discussion}
 Isochronous systems occur in various physical situations and have been widely investigated in the literature with the prototypical systems being the harmonic and isotonic oscillators. In the present work, we discussed the generalization of the isotonic oscillator in the framework of relativistic quantum mechanics by proposing the Dirac-isotonic oscillator in \((1+1)\) and \((2+1)\) dimensions, making use of supersymmetric quantum mechanics. This generalized system in \((1+1)\) dimensions reduces to the (shifted) isotonic oscillator in the non-relativistic limit, whereas the Dirac-isotonic oscillator in \((2+1)\) dimensions maps to an isotonic oscillator with strong spin-orbit coupling and an added centrifugal term that couples with the orbital angular momentum leading to a shift in the quantum number \(m_l\). In both \((1+1)\) and \((2+1)\) dimensions, the systems are exactly solvable and their spectra (for \(E^2\)) are equispaced. 
 
 \vspace{2mm}

 Let us conclude this study by pointing out a possible generalization of the connection \cite{qpt1} between the Dirac oscillator in \((2+1)\) dimensions and the Jaynes-Cummings/anti-Jaynes-Cummings model \cite{opt,opt1,opt2}. In the preceding sections, we made extensive use of the framework of supersymmetric quantum mechanics to introduce Dirac-isotonic oscillators in \((1+1)\) and \((2+1)\) dimensions. The essential idea was to choose suitable superpotential functions. Generally, one can define the supercharge operators
 \begin{equation}
\mathcal{A}^\dagger = \frac{P^\dagger + i \mathcal{W}^\dagger(\mathbf{r})}{\sqrt{\delta}}, \quad \quad \mathcal{A} = \frac{P - i \mathcal{W}(\mathbf{r})}{\sqrt{\delta}},
\end{equation} where \(P = p_x + i p_y\), \(\mathcal{W}(\mathbf{r}) = W_x(\mathbf{r}) + i W_y(\mathbf{r})\), and \(\delta\) is a positive constant. It then follows that
\begin{widetext}
\begin{equation}
 \mathcal{A}^\dagger \mathcal{A} \delta= p_x^2 + p_y^2 + W_x^2 + W_y^2 + 2(W_y p_x - W_x p_y) - i ((p_x W_x) + (p_y W_y)) + ((p_xW_y) - (p_y W_x)),
\end{equation}
while 
\begin{equation}
\mathcal{A}\mathcal{A}^\dagger \delta = p_x^2 + p_y^2 + W_x^2 + W_y^2 + 2(W_y p_x - W_x p_y) + i ((p_x W_x) + (p_y W_y)) + ((p_xW_y) - (p_y W_x)).
\end{equation}
\end{widetext}
Naturally then, one gets the commutator \([\mathcal{A},\mathcal{A}^\dagger] = 2i \delta^{-1} ((p_x W_x) + (p_y W_y))\). Taking \(W_x = m \omega x\) and \(W_y = m \omega y\) as taken for the \((2+1)\)-dimensional Dirac oscillator and upon picking \(\delta = 4 m \omega\), one gets \([\mathcal{A},\mathcal{A}^\dagger] = 1\), the familiar oscillator algebra. Generally, with nonlinear superpotentials such as that giving the Dirac-isotonic oscillator, one has \([\mathcal{A},\mathcal{A}^\dagger] \neq 1\). Nevertheless, employing these operators one can express the Dirac Hamiltonian as
 \begin{equation}\label{2dgenmatrixHnewAexp}
H =
\begin{pmatrix}
mc^2 & c \sqrt{\delta} \mathcal{A}^\dagger \\
c\sqrt{\delta} \mathcal{A} & -mc^2
\end{pmatrix} .
\end{equation}
Since the mapping between the \((2+1)\)-dimensional Dirac oscillator and the Jaynes-Cummings/anti-Jaynes-Cummings model \cite{opt,opt1,opt2} is well known \cite{qpt1} (see also, \cite{diracB}), the above-mentioned generalized framework directly allows one to introduce a generalized Jaynes-Cummings/anti-Jaynes-Cummings model with creation-annihilation operators being replaced by the supercharges. To this end, introducing the spin-projection operators \(\sigma^\pm = \frac{1}{2} (\sigma_x \pm \sigma_y)\), it is simple to show that the Hamiltonian matrix (\ref{2dgenmatrixHnewAexp}) is equivalent to the following form:
 \begin{equation}\label{HAJC}
 H = g (\sigma^- \mathcal{A} + \sigma^+ \mathcal{A}^\dagger) + \sigma_z mc^2,
 \end{equation} where the parameter describing the spin-supercharge coupling is given by \(g = c \sqrt{\delta}\). Notice that while the standard anti-Jaynes-Cummings Hamiltonian \cite{opt} describes the coupling between a two-level system and a single-mode oscillator, the supersymmetric generalization presented above describes coupling between the spin operators (two-level) and the supercharge operators. It would be exciting to explore the possibility of a relationship between the Dirac-isotonic oscillator and optical systems. Let us keep this issue for future work. 
\\

\textbf{Acknowledgements:} We are thankful to Bijan Bagchi for several stimulating discussions. A.G. is grateful to Akash Sinha for discussions and also thanks the Ministry of Education, Government of India for financial support in the form of a Prime Minister's Research Fellowship  (ID: 1200454). A.G. is grateful to the Department of Physics, Banaras Hindu University for hospitality where this work was initiated. B.P.M. acknowledges the incentive research grant for faculty under the IoE Scheme (IoE/Incentive/2021-22/32253) of the Banaras Hindu University.

\end{document}